\documentclass[conference,compsoc]{IEEEtran}
\IEEEoverridecommandlockouts

\usepackage[utf8]{inputenc}

\ifCLASSOPTIONcompsoc
  \usepackage[nocompress]{cite}
\else
  \usepackage{cite}
\fi
\ifCLASSINFOpdf
  \usepackage[pdftex]{graphicx}
  \graphicspath{{.}}
  \DeclareGraphicsExtensions{.pdf,.jpeg,.png}
\else
\fi
\usepackage{fancyhdr}

{
\fancyhf{}
\fancyfoot[L]{A modified version of this manuscript will be published in the Proceedings of the IEEE DataCom 2016 \textcopyright 2016 IEEE}
\fancyfoot[C]{}
\fancyfoot[R]{}

}

\begin{document}
%
\title{Generation of the Single Precision BLAS library for the Parallella platform, with Epiphany co-processor acceleration, using the BLIS framework}

\author{\IEEEauthorblockN{Miguel Tasende}
\IEEEauthorblockA{Research \& Development Department\\
Antel\\
Montevideo, Uruguay\\
Email: mtasendebracco@antel.com.uy}
}


%


\IEEEpubid{\makebox[\columnwidth]{\copyright~2016 IEEE}
\hspace{\columnsep}\makebox[\columnwidth]{A modified version of this manuscript will be published in the Proceedings of the IEEE DataCom 2016}}

\maketitle

\begin{abstract}
The Parallella is a hybrid computing platform that came into existence as the result of a Kickstarter project by Adapteva. It is composed of the high performance, energy-efficient, manycore architecture, Epiphany chip (used as co-processor) and one Zynq-7000 series chip, which normally runs a regular Linux OS version, serves as the main processor, and implements ``glue logic'' in its internal FPGA to communicate with the many interfaces in the Parallella. In this paper an Epiphany-accelerated BLAS library for the Parallella platform was created (which could be suitable, also, for similar hybrid platforms that include the Epiphany chip as a coprocessor). For the actual instantiation of the BLAS, the BLIS framework was used. There have been previous implementations of Matrix-Matrix multiplication, on this platform, that achieved very good performances inside the Epiphany chip (up to 85\% of peak), but not so good ones for the complete Parallella platform (due to inter-chip data transfer bandwidth limitations). The main purpose of this work was to get closer to practical Linear Algebra aplications for the entire Parallella platform, with scientific computing in view.\\
\\

A modified version of this manuscript will be published in the Proceedings of the IEEE DataCom 2016 \textcopyright 2016 IEEE
\end{abstract}


%
\IEEEpeerreviewmaketitle

\section{Introduction}
In recent times there has been interest in the use of hybrid platforms (mostly CPUs with GPUs or Manycore accelerators) for scientific computation in large clusters. On the other hand RISC-based clusters, and ARM-based ones in particular, are also of interest, among other things, because of the low power consumption that is achievable on those architectures, and because new consumer products have made them ubiquitous (smartphones, tablets, etc.), lowering their cost. It is possible to think that the same way the consumer PC ``explosion'' gave many cheap hardware for use in modern HPC clusters (directly or indirectly), the ``mobile'' products could lead to new improvements in HPC infrastructure.\\
The Parallella platform \cite{kickstarting} has both: it's a hybrid platform based on an ARM CPU, and a manycore RISC device as a co-processor (the Epiphany) \cite{parallellaReference}.
In this work the real and practical possibilities of the Parallella platform for Scientific Computing are explored. To have a starting point, the Linpack benchmark was chosen to be run on a cluster of Parallella nodes, but it was found that there was no (Epiphany accelerated) BLAS implementation for the platform. Therefore, a BLAS library was ``instantiated'' with the BLIS framework \cite{BLIS1}, after writing an Epiphany accelerated sgemm\footnote{sgemm: ``Single Precision, General Matrix Multiplication''} micro-kernel for it.\\
The idea for the micro-kernel was to use a ``SUMMA-like'' algorithm \cite{summa}, that could improve the performance over current implementations (that use Cannon's \cite{cannon}). The achieved results, for the Matrix-Matrix Multiplication performance, were the best for this platform that are presently known to the author \cite{varghese}\cite{ysapir}\cite{Ross2016} (if the host processing and off-chip data transfer is taken into account).\\

In the following sections a very brief overview of the Parallella Board is given and then the current solution implementation, for instatiating the BLAS library, is explained. It was followed a ``top-bottom'' approach, in which the highest level parts of the system are explained first and the low-level parts later. In section 4 the results for a number of benchmarks are shown and in section 5 the conclusions and future work are stated.

\hfill mds
 
\hfill March 29, 2016

\section{The Parallella board}
\begin{figure}[!h]
\centering
\includegraphics[width=3.5in]{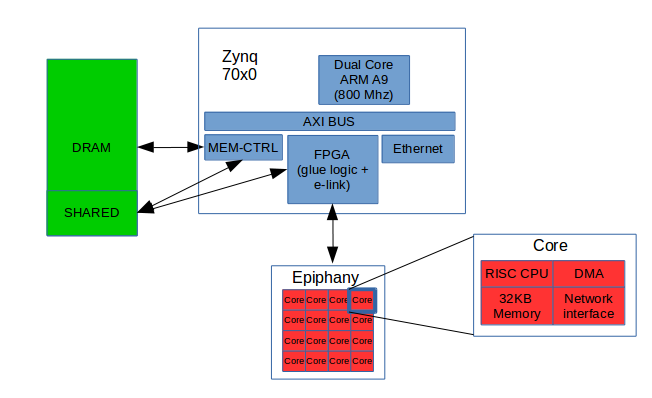}
\caption{The Parallella Architecture.}
\label{fig_parallellaArch}
\end{figure}

\IEEEpubidadjcol
The Parallella board\cite{parallellaReference} has one Zynq 7010 or 7020 chip acting as ``the host processor'', one Epiphany chip acting as a ``co-processor'', and a 1GB DRAM chip, of which 32MB are accessible to both the host and coprocessor (shared DRAM). It also contains many interfaces, like Ethernet, USB, a slot for an SD card, etc., to communicate with other hardware.\\
The Zynq SoC\cite{Zynq7000} can be basically thought as a dual-core ARM Cortex-A9 CPU, with an FPGA embedded, and many on-chip interfaces. The FPGA is used to implement the ``e-link'' that is needed to communicate with the Epiphany chip. That communications interface also allows the Epiphany chip to access the shared portion of RAM (32MB).\\
The Epiphany chip\cite{epiphanyArchReference} consists of a 2D array of cores (``eCores'') connected by a mesh Network-on-chip. Each core contains a RISC CPU, a DMA engine, 32 KB of local memory and a Network interface (see figure \ref{fig_parallellaArch}).\\
To program the Parallella architecture there are different options. The one chosen here was the eSDK\cite{epiphanySDKReference} provided by Adapteva, which consists of a series C functions that allow the communication between a host and the Epiphany SoC (grouped in the ``e-hal'' library), and between eCores within the Epiphany (grouped in the ``e-lib'' library). Among other things, the host can load programs to individual eCores, write and read the eCores' local memory, and interrupt them. The ``standard'' model for accelerating a normal C function running on the host would be:
\begin{enumerate}
\item The host runs initialization code, and defines workgroups
\item The host loads kernel programs to the workgroups
\item The host sends the input data (either directly or through the shared RAM)
\item The host signals the workgroups to start
\item The coprocessor gets the input data (from shared RAM or local memory) and processes it
\item The coprocessor sends the output data (to shared RAM or local memory)
\item The coprocessor signals the end of the calculations
\item The host gets the results and continues with the execution of the main process
\end{enumerate}
It is important to note that the Epiphany kernels can be written in C (although it is not always possible to achieve the best performance in that language, and some assembly code may be needed).

\section{Software Architecture}
\subsection{BLIS}
BLIS is a portable software framework for instantiating high-performance BLAS-like dense linear algebra libraries\cite{BLIS1}. When invoked, it generates a new BLAS-like API that its creators made to improve the old BLAS library, but also generates the classic FORTRAN BLAS library, and allows to write custom C micro-kernels to accelerate the resulting BLAS functions. That is the use that was given on this work. A micro-kernel was written to accelerate the ``sgemm'' function by offloading the main calculations to the Epiphany coprocessor. When a BLAS user (may be any scientific software, or library like LAPACK, ScaLAPACK, etc.) calls the ``sgemm'' function, the BLIS code divides the input and output matrices conveniently and sends small predefined multiplications to be performed by the micro-kernel.

The custom micro-kernel, after performing some initialization tasks, calls the Epiphany to do the heavy part of the calculations, does some post-processing, and then returns the partial results to the bigger sgemm function.

The problem that the (BLIS-generated) sgemm should solve is: given $A \in M_{M\times K}$, $B \in M_{K\times N}$, $C_{in} \in M_{M\times N}$, then calculate $C_{out} = \alpha A\cdot B + \beta C_{in}$, possibly transposing some of the matrices, and taking into account their correct representation in memory (leading dimensions), where $M$, $N$ and $K$ are arbitrary.

\subsection{A Separate Linux process}
The first task of the micro-kernel\footnote{The ``micro-kernel'' is part of the host process. It is called ``micro-kernel'' by using the BLIS nomenclature (it is the kernel of the bigger ``sgemm'' function inside the generated BLAS). It is not to be confused with the ``Epìphany kernel'' which runs in the coprocessor.} is to initialize the communications with the Epiphany chip and reset the system (or parts of it), it defines the shared blocks of RAM, then defines the workgroups (one on this case), loads the coprocessor kernels (only one in this case), and starts the workgroups. When the coprocessor kernel finishes the calculations (possibly many ``coprocessor tasks''), the micro-kernel has to free the allocated shared RAM, and close the connection to the coprocessor.\\
All that operations, on one hand, take a lot of time, and on the other, it was found that some of the ``initialize/finalize'' functions of the eSDK had technical problems when called many times by the same process (in this case the BLAS process calling the micro-kernel many times would be doing so). The solution for those problems was to place the initialization and finalization code in an entirely different process that runs as a ``linux service'' (in the research version is just a different process but it could be easily converted to a Linux daemon). With that solution in mind, we will now have a Host-Coprocessor shared RAM, and also a Host-Host shared RAM. They will be called the HC-RAM and HH-RAM, respectively. Of course if it was possible to use the same space for both communications some time could be saved, but that was not yet implemented on this work.

The basic scheme is this: The BLAS sgemm function calls the micro-kernel, the micro-kernel sends its input data to a predefined place in the HH-RAM (using POSIX Shared Memory tools) and passes the control to the ``service process'' (with a semaphore). The service process has already intialized all the necessary structures, has established communication with the coprocessor, and loaded the kernel, which is waiting for a signal to start processing. It gets the data from the HH-RAM, and runs the ``sgemm inner micro-kernel'' that is explained below.

\subsection{sgemm inner micro-kernel}
\begin{figure}[!h]
\centering
\includegraphics[width=3.5in]{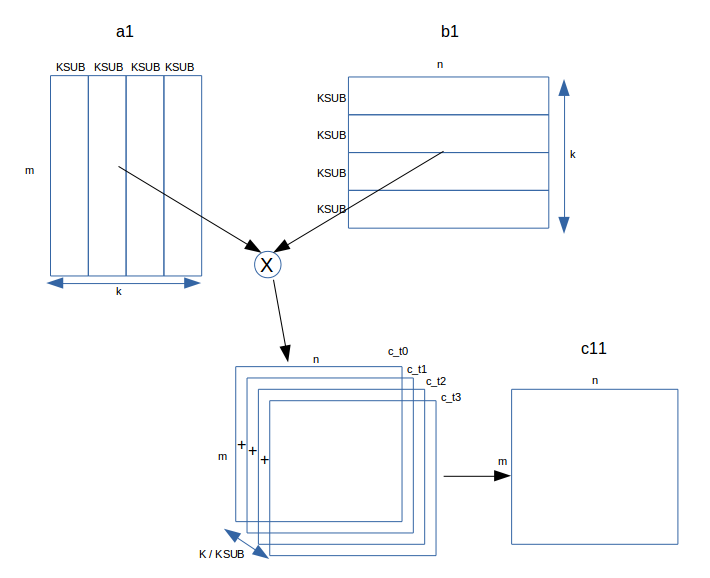}
\caption{The ``sgemm inner micro-kernel'' host algorithm. Each $c_{ti}$ is calculated by an ``Epiphany Task''.}
\label{fig_BLISMicroKernel}
\end{figure}

Some variables are to be defined.
\begin{itemize}
\item $ir$ is the ratio of the input loading and preprocessing host time to the total time of the sgemm inner micro-kernel.
\item $or$ is the ratio of the postprocessing host time to the total time of the sgemm inner micro-kernel.
\end{itemize}

The ``sgemm inner micro-kernel'' is the name that will be given to the host portion of the code that is run from within the service process (after the transfer of data and signaling by the main process). It includes the outer part of the multiplication algorithm that is performed by the Epiphany coprocessor. It follows a ``SUMMA-like'' scheme, and then it does some post-processing.\\

The problem that the sgemm inner micro-kernel should solve is: given $a1 \in M_{m\times K}$, $b1 \in M_{K\times n}$, $c_{in} \in M_{m\times n}$, then calculate $c_{out} = \alpha a1\cdot b1 + \beta c_{in}$, where $m$ and $n$ are fixed, known and configurable, $K$ is arbitrary, and the leading dimensions of all matrices are known ($a1$ is column-major stored, $b1$ is row-major stored and $c_{in}$, $c_{out}$ are column-major stored). On the other hand, the column-strides and row-strides of the input and output matrices are arbitrary and given as input to the kernel (it has to handle the different possible strides).

The process of calculation of the micro-kernel is as follows. The input matrices ($a1$, $b1$) are divided in blocks of $KSUB$ columns and rows respectively (they are divided in the ``k dimension''). The main loop iterates on those blocks, sending one $(m \times KSUB)$-size block from input $a1$, and one $(KSUB \times n)$-size block from input $b1$ on each iteration (see figure \ref{fig_BLISMicroKernel}). Those input blocks for an ``Epiphany Task'' will be called $a_{ti}$ and $b_{ti}$, respectively. The ``Epiphany Task'' takes care of performing the outer product of each column of $a_{ti}$ with each row of $b_{ti}$ that the micro-kernel has sent, and it performs a partial sum of those products. The result is the partial result matrix (for task $i$): $c_{ti}$. Each of those $\frac{K}{KSUB}$ partial results (that are $(m\times n)$ in size) can then be summed by the (host) sgemm inner micro-kernel, or can be accumulated in the coprocessor local memory, depending on the implementation (see figure \ref{fig_BLISMicroKernel}). After that, the micro-kernel multiplies the resulting matrix by $\alpha$ and adds $\beta \cdot c_{in}$, to produce the sgemm micro-kernel final result. It stores it in the HH-RAM and signals the main process (sgemm outer micro-kernel) that the calculation is done.\\
The data exchange between host and co-processor is done via the shared RAM (HC-RAM). The process of sending the inputs is interleaved with the Epiphany Task (while the task is executing on the co-processor the host is sending the next KSUB-block to the HC-RAM). To achieve that interleaving there are two buffers reserved for each input block, and a shared control variable (``selector'') that tells the co-processor in which buffer the input is, for the current iteration.

There is another shared control variable (``command'') that tells the coprocessor what to do in the current iteration:

\begin{itemize}
\item $command=0$) Clear the inner buffers (initialization) and proceed with one Epiphany Task. Don't send the results back.\\
\item $command=1$) Proceed with one Epiphany Task. Don't clear buffers, or send results.\\
\item $command=2$) Proceed with one Epiphany Task and send the results back. Don't clear buffers.\\
\item $command=3$) ``There will be a unique iteration'': Clear the buffers, do one Epiphany Task, and send the results back.\\
\end{itemize}

Using the ``command'' variable, the host micro-kernel can tell the coprocessor to do the initialization steps only once, then accumulate the results of many KSUB-blocks, and in the last iteration send the final result back. Thus a lot of time is saved (most importantly the time needed to ``send the results back''). When that scheme is used, the algorithm will be called ``An Accumulator'', and it reduces the output times and postprocessing ratio ($or$) to near zero as K is made larger. The dissadvantage of accumulating is that the results (of $m\times n$ size) must be stored fully in the local memory, and that limits the maximum possible size of $m$ and $n$. $m$ and/or $n$ increases are needed to reduce the input time ratio ($ir$). So, a clear compromise exists between improving the $or$ and the $ir$ ratios.

\subsection{Epiphany kernel}
\begin{figure}[!h]
\centering
\includegraphics[width=2.5in]{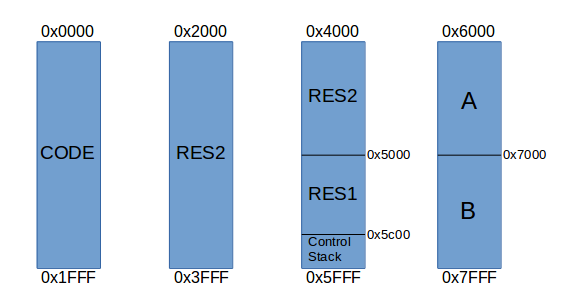}
\caption{Local memory mapping for one core. A and B are the inputs. RES2 is a buffer to store the entire result part that corresponds to this core, and is also used as one of the temporary communications buffers in the ``K Iteration'', while RES1 is used as the other temporary communications buffer. The stack and some control variables have a reserved region. The first bank is used by the kernel's code.}
\label{fig_memoryMap}
\end{figure}

Due to the memory restrictions it is very important to organize the code and buffers in the local memory. In figure \ref{fig_memoryMap} it is shown the local memory map, for one core, in this implementation.

\subsubsection{Epiphany Task}

The outer layer of the Epiphany kernel will be called an ``Epiphany Task''\footnote{Some definitions: CORES is the number of cores in the Epiphany chip. KSUB is the number of columns of a1 and rows of b1 that are sent to the Epiphany chip on each Epiphany Task. NSUB is the number of columns of one subMatmul result.}. Again, the algorithm is ``SUMMA-like'' \cite{summa}. The input is divided between the cores in blocks of $(m\times \frac{KSUB}{CORES})$ size for $a_{ti}$ and $(\frac{KSUB}{CORES} \times n)$ size for $b_{ti}$ (those block will be called $a_{ti-cj}$ and $b_{ti-cj}$). Each core will calculate the correspondig outer products and sum over $\frac{KSUB}{CORES}$ of them to obtain a partial result ($c_{ti-cj}$) that, in turn, will be summed with the partial results of the other cores (resulting $c_{ti}$). It is important to note how this inter-core summing is achieved.\\
Each core is the ``owner'' of one part of the final matrix result, which it will store after the Epiphany Task is run. The partition could be made arbitrarily, but on this implementation it was chosen to divide the results matrix in blocks of $\frac{n}{CORES}$ columns, each. That was done in order to make the reorganization of the output matrix easier (as it is stored column-major). It also allows for the ``b-streaming'' implementations, in which the input matrix $a_{ti-cj}$ is totally stored in local memory, but the $b_{ti-cj}$ input matrix is retrieved ``as needed'' by the coprocessor (in blocks of $NSUB \cdot CORES$), as will be explained later.\\
In this ``input storage - output storage'' scheme it can be readily seen that the input needed to calculate the output results of one core lies scattered around the other cores. Usually, the solution would be to move partial input data within the cores, as moving results would be more costly, but on this case (due to some Epiphany special characteristics) the implementation moves the partial results instead. The idea is to make use of the fact that the Epiphany cores can do one ``multiply-add'' and one ``store into another core's memory'' on the same clock cycle, so the results inter-core movement can be done ``for free'', which can't be done for the inputs. An inter-core pipeline (figure \ref{fig_epiphanyPipeline}) was designed to move those intermediate results, as will be explained below. Resuming, the core that is responsable of a certain calculation is not necessarily responsable of the final storage of it, and the storage scheme can be chosen arbitrarily (on this implementation divided by column blocks).
\begin{figure}[!h]
\centering
\includegraphics[width=3.5in]{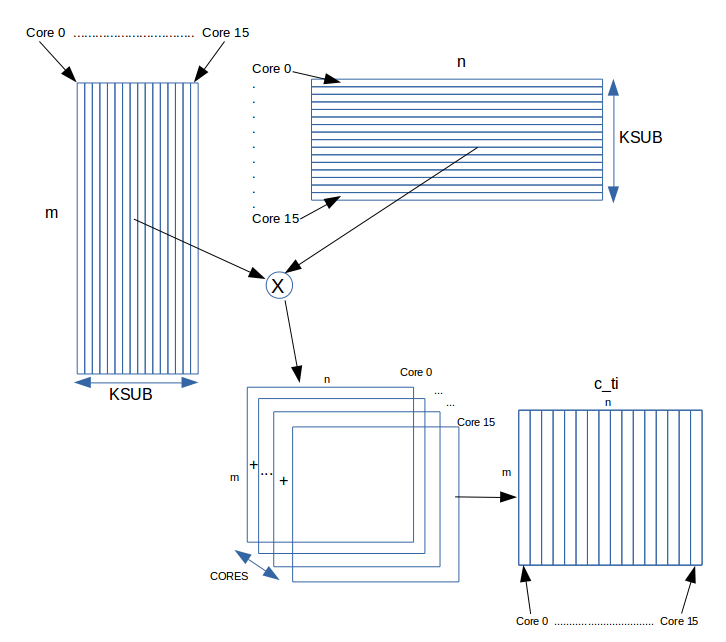}
\caption{One Epiphany Task.}
\label{fig_epiphanyTask}
\end{figure}

\subsubsection{Epiphany Column Iteration}
\begin{figure}[!h]
\centering
\includegraphics[width=3.5in]{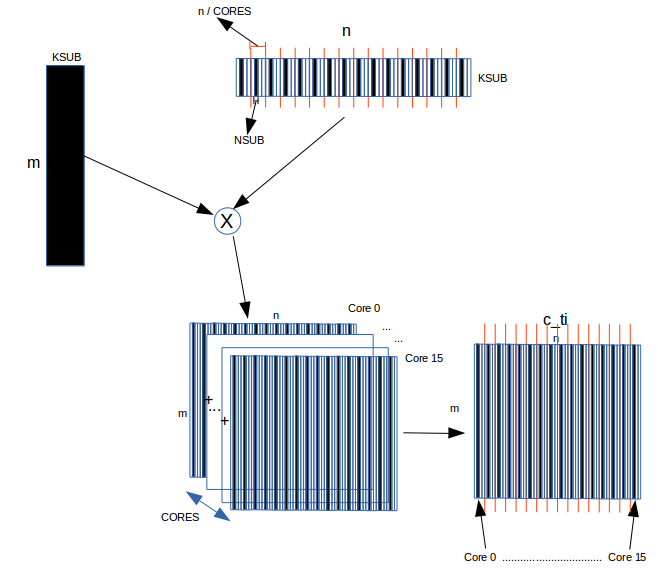}
\caption{One Epiphany Column Iteration. The sections of the input and output that take part in the process were colored in black.}
\label{fig_epiphanyColumnIteration}
\end{figure}

The $b_{ti}$ input matrix to the Epiphany Task is divided in $CORES$ blocks of size $KSUB\times \frac{n}{CORES}$, that correspond (ultimately) to the cores' output storage blocks. Furthermore, each of those blocks are divided in blocks of size $KSUB\times NSUB$. It will be called an ``Epiphany Column Interation'' (see figure \ref{fig_epiphanyColumnIteration}) to the calculation of $CORES$ non-adjacent blocks of size $m\times NSUB$, of the final Epiphany Task result matrix. Each Epiphany Column Iteration consists of $CORES$ ``Epiphany K Iterations'', that will be described in the next subsection. On an Epiphany Column Iteration, each core calculates $CORES$ partial results of size $m\times NSUB$ and in the end stores a final result block of size $m\times NSUB$ (that means that, in total, there are $CORES$ of those final result blocks calculated). After $\frac{n}{NSUB}$ Epiphany Column Iterations, the Epiphany Task is completed.

\subsubsection{Epiphany K Iteration}
\begin{figure}[!h]
\centering
\includegraphics[width=2.0in]{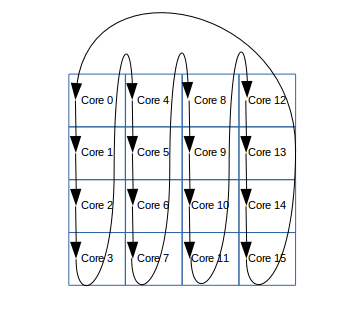}
\caption{The Epiphany pipeline.}
\label{fig_epiphanyPipeline}
\end{figure}

On each ``Epiphany Column Iteration'' is divided into $CORES$ ``Epiphany K Iterations''. Each Epiphany K Iteration a partial result block of size $m\times NSUB$ is calculated by each core, and sent to the next core in the defined pipeline (figure \ref{fig_epiphanyPipeline}) to be accumulated with other partial results. The identity of the next core is fixed, but the position, in the final results matrix, of the block that is calculated depends on the current  iteration number as much as on the id of the core that performs the calculation.

\begin{figure}[!h]
\centering
\includegraphics[width=3.5in]{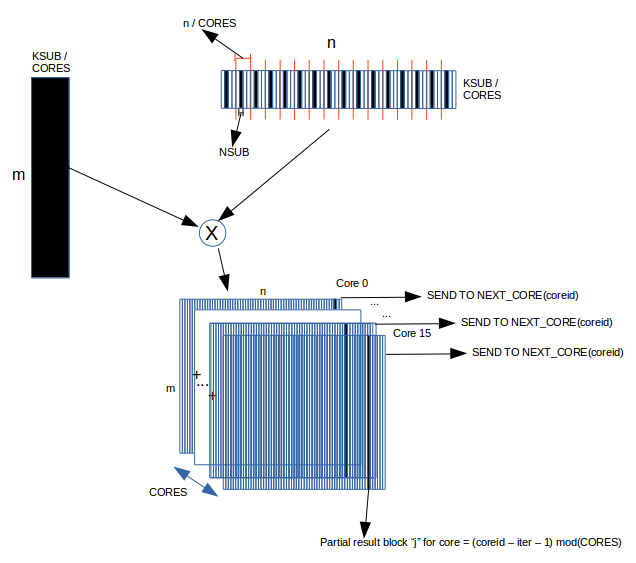}
\caption{One Epiphany K Iteration. The regions of the input and output that take part in the process were colored in black.}
\label{fig_EpiphanyKIteration}
\end{figure}

On every K Iteration, a partial block (corresponding to a partial sum of blocks in the current ``Column Iteration'' position), that will ultimately end in the core number $(ownCoreid - iter_k - 1) mod(CORES)$, is sent to the next core. Thus, after $CORES$ iterations every core has its own results block.

As an example, on iteration zero, core 0, calculates one partial block corresponding to core 15, and sends it to core 1. On iteration two it will calculate a block corresponding to core 14 and sends it to core 1, and so on. On iteration $CORES-1$, it calculates a partial block corresponding to core 1, and sends it to core 1, and in the final iteration it calculates it's own correspondig block and sends it to a different destination depending on the value of the ``command'' variable. If the command asks to send the results out, it will copy them to the HC-RAM. Otherwise, if the results are to be accumulated, core 0 sends its results to core 1 as in previous iterations (when new input data arrives this will correctly accumulate the new results with the old).

For sending and receiving the partial results, two buffers are defined and are interchanged on even and odd K iterations. One of the buffers has a size to hold the entire final result ($m\times n$), but is used, on each Epiphany Column Iteration, in blocks of size $m\times NSUB$ (it doesn't change in the K Iterations loop). There is a second, fixed, buffer of size $m\times NSUB$. On each K Iteration, one of the buffers is the holder of the ``previous accumulated result'' and the other is used to store the current result (in the next core). The initial buffer is defined so that in the last K Iteration the results are in the (big, final results) RES2 buffer.\\
Before and after every K Iteration a barrier is used to synchronize the cores.

\subsubsection{subMatmul}
The function ``subMatmul'' could be thought as ``the single-core version of the Epiphany K Iteration''. It is just a single-core matrix-matrix multiplication function, that accepts inputs of size $m\times \frac{KSUB}{CORES}$ for $a$, and $\frac{KSUB}{CORES}\times n$ for $b$, and outputs the resultant $m\times NSUB$ product matrix. This function was initially implemented in C language, but as it became clear that it was the critical function in the kernel, it was then implemented in assembly language. The implementation was strongly based on that of the previous work \cite{varghese}, which achieved on-chip performances close to $85\%$ of the peak. That implementation is based on a macro ``doMult'' that basically multiplies one scalar with an array of size 32. It makes use of the Epiphany core's special features to achieve great performances (see section VII of \cite{varghese} for details).\\
The assembly version has fixed input and output sizes: $a\in M_{192\times 4}$, $b\in M_{4\times 4}$, $C_{in}, C_{out}\in M_{192\times 4}$. The previous result and next result pointers, are passed as parameters.\\
In this implementation the doMult macro was repeated 4 times (as matrices are of size 4 in the ``k'' dimension), which means that the partial results will be accumulated 4 times in the internal registers, before sending them back to memory. As the length of one ``doMult result vector'' is 32, a loop that repeats the process 6 times was necessary, to calculate a complete 192 column. After that inner loop, another outer loop iterates on the ($NSUB=4$) $b$ columns and repeats the process.

\begin{figure}[!h]
\centering
\includegraphics[width=2.5in]{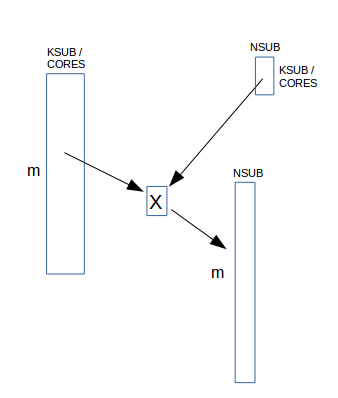}
\caption{Scheme of the subMatmul function.}
\label{fig_subMatmul}
\end{figure}

\section{Results}

\subsection{Custom Tests}
All the processing times (in these tests) were measured in the host side with functions from the ``\verb+time.h+'' C library. In the case of the kernel called from the same OS process, the times where measured with the ``clock()'' function. In the case of the kernel called from a different process (which is the actual implementation that was compiled for BLIS), the times are measured with calls to \verb+clock_gettime(CLOCK_MONOTONIC,&time)+. That was necessary because the ``clock()'' function would give results only for the main process. The results of both measurements can be seen in tables \ref{table_custom1} and \ref{table_custom2}.

\begin{table}[!h]
\renewcommand{\arraystretch}{1.3}
\caption{Custom Tests results for the sgemm kernel called from the same process (m=192, n=256, K=4096).}
\label{table_custom1}
\centering
\begin{tabular}{|p{0.2\textwidth}|c|c|c|c|}
\hline
Description  & Time (s) & \% & GFLOPS/s \\
\hline
Host reference code  & 3.778169 & 100  & 0.107\\
\hline
Input loading and host preprocessing (*) & 0.094648 & 82.9 & -\\
\hline
Coprocessor work (*) & 0.105652 & 92.6 & -\\
\hline
Host data retreiving and postprocessing & 0.005272 & 4.6 & -\\
\hline
Total sgemm $\mu$-kernel & 0.114114 & 100 & 3.529\\
\hline
Mean Relative Error & \multicolumn{3}{|c|}{8.73e-08}\\
\hline
Maximum Relative Error & \multicolumn{3}{|c|}{5.83e-07}\\
\hline
\multicolumn{4}{p{0.45\textwidth}}{(*) Input loading and coprocessor work are done in parallel, which explains that the sum of the percentaje column, for the sgemm $\mu$-kernel, is larger than 100}\\
\end{tabular}
\end{table}

\begin{table}[!h]
\renewcommand{\arraystretch}{1.3}
\caption{Custom Tests results for the sgemm kernel called from a different process (m=192, n=256, K=4096).}
\label{table_custom2}
\centering
\begin{tabular}{|p{0.2\textwidth}|c|c|c|c|}
\hline
Description  & Time (s) & \% & GFLOPS/s \\
\hline
Host reference code  & 3.776418 & 100  & 0.107\\
\hline
Total sgemm $\mu$-kernel & 0.158303 & 100 & 2.543\\
\hline
Mean Relative Error & \multicolumn{3}{|c|}{8.73e-08}\\
\hline
Maximum Relative Error & \multicolumn{3}{|c|}{5.83e-07}\\
\hline
\end{tabular}
\end{table}

\subsection{BLIS Tests}
After the ``custom'' tests, the BLIS (and BLAS) library was compiled with the micro-kernel, and the BLIS standard tests were run. The micro-kernel used is the one that calls a different OS process to calculate the results. As can be seen in table \ref{table_BLISk1} the results are very similar to those of the ``custom'' tests.

\begin{table}[!h]
\renewcommand{\arraystretch}{1.3}
\caption{BLIS sgemm kernel results (m=192, n=256, K=4096)}
\label{table_BLISk1}
\centering
\begin{tabular}{|p{0.2\textwidth}|c|c|c|}
\hline
\verb+blis_<dt><op>+ \verb+_<params>_<stor>+ & GFLOPS & residue\\
\hline
\verb+blis_dgemm_nn_ccc+  & 2.630 & 1.18e-07\\
\hline
\end{tabular}
\end{table}

In table \ref{table_BLIS1} the tests (from the ``BLIS testsuite'') for the whole sgemm function, are shown with $m=n=K=4096$. It can be seen that the performance penalty, with respect to the kernel performance, is not too big.

\begin{table}[!h]
\renewcommand{\arraystretch}{1.3}
\caption{BLIS sgemm results (m=4096, n=4096, K=4096)}
\label{table_BLIS1}
\centering
\begin{tabular}{|p{0.2\textwidth}|c|c|c|}
\hline
\verb+blis_<dt><op>+ \verb+_<params>_<stor>+ & GFLOPS & residue\\
\hline
\verb+blis_sgemm_nn_ccc+  & 2.381 & 4.52e-07\\
\hline
\verb+blis_sgemm_nc_ccc+  & 2.381 & 4.79e-07\\
\hline
\verb+blis_sgemm_nt_ccc+  & 2.455 & 4.77e-07\\
\hline
\verb+blis_sgemm_nh_ccc+  & 2.456 & 4.65e-07\\
\hline
\verb+blis_sgemm_cn_ccc+  & 2.381 & 4.69e-07\\
\hline
\verb+blis_sgemm_cc_ccc+  & 2.381 & 4.75e-07\\
\hline
\verb+blis_sgemm_ct_ccc+  & 2.455 & 4.67e-07\\
\hline
\verb+blis_sgemm_ch_ccc+  & 2.455 & 4.59e-07\\
\hline
\verb+blis_sgemm_tn_ccc+  & 2.034 & 4.50e-07\\
\hline
\verb+blis_sgemm_tc_ccc+  & 2.036 & 4.64e-07\\
\hline
\verb+blis_sgemm_tt_ccc+  & 2.090 & 4.55e-07\\
\hline
\verb+blis_sgemm_th_ccc+  & 2.094 & 4.89e-07\\
\hline
\verb+blis_sgemm_hn_ccc+  & 2.035 & 4.67e-07\\
\hline
\verb+blis_sgemm_hc_ccc+  & 2.037 & 4.69e-07\\
\hline
\verb+blis_sgemm_ht_ccc+  & 2.090 & 4.69e-07\\
\hline
\verb+blis_sgemm_hh_ccc+  & 2.094 & 4.63e-07\\
\hline
\multicolumn{3}{p{0.45\textwidth}}{dt=data type, op=operation, params=[(n)o-transpose, (t)ranspose, (c)onjugate, (h)ermitian-transpose]. The ``c'' and ``h'' options are the same as ``n'' and ``t'' respectively, as the tests are for real values, in this case.}
\end{tabular}
\end{table}

As the version of the HPL Linpack Benchmark code that was readily available to the author was intended for use with Double Precision, and with the goal of making a first test, that would further establish the correctness and robustness of the solution, a ``dgemm'' kernel was implemented, for the BLIS framework which, in fact, sends the data to the ``sgemm inner kernel'' to do the calculations (downcasting the inputs, and upcasting the outputs). The precision of the results is, therefore, expected to be close to that of Single Precision. It was a workaround to be able to reuse the already available HPL code. In the process, some performance was lost, as can be seen in table \ref{table_BLISk2}. That version was called the ``false dgemm''.

\begin{table}[!h]
\renewcommand{\arraystretch}{1.3}
\caption{BLIS ``false dgemm'' kernel results (m=192, n=256, K=4096)}
\label{table_BLISk2}
\centering
\begin{tabular}{|p{0.2\textwidth}|c|c|c|}
\hline
\verb+blis_<dt><op>+ \verb+_<params>_<stor>+ & GFLOPS & residue\\
\hline
\verb+blis_dgemm_nn_ccc+  & 2.073 & 9.33e-09\\
\hline
\end{tabular}
\end{table}

In table \ref{table_BLIS2} the results for the whole ``false dgemm'' function are shown.

\begin{table}[!h]
\renewcommand{\arraystretch}{1.3}
\caption{BLIS ``false dgemm'' results (m=4096, n=4096, K=4096)}
\label{table_BLIS2}
\centering
\begin{tabular}{|p{0.2\textwidth}|c|c|c|}
\hline
\verb+blis_<dt><op>+ \verb+_<params>_<stor>+ & GFLOPS & residue\\
\hline
\verb+blis_dgemm_nn_ccc+  & 1.785 & 1.30e-08\\
\hline
\verb+blis_dgemm_nc_ccc+  & 1.785 & 1.28e-08\\
\hline
\verb+blis_dgemm_nt_ccc+  & 1.829 & 1.32e-08\\
\hline
\verb+blis_dgemm_nh_ccc+  & 1.828 & 1.28e-08\\
\hline
\verb+blis_dgemm_cn_ccc+  & 1.784 & 1.30e-08\\
\hline
\verb+blis_dgemm_cc_ccc+  & 1.783 & 1.29e-08\\
\hline
\verb+blis_dgemm_ct_ccc+  & 1.828 & 1.28e-08\\
\hline
\verb+blis_dgemm_ch_ccc+  & 1.828 & 1.29e-08\\
\hline
\verb+blis_dgemm_tn_ccc+  & 1.580 & 1.27e-08\\
\hline
\verb+blis_dgemm_tc_ccc+  & 1.578 & 1.29e-08\\
\hline
\verb+blis_dgemm_tt_ccc+  & 1.613 & 1.28e-08\\
\hline
\verb+blis_dgemm_th_ccc+  & 1.611 & 1.26e-08\\
\hline
\verb+blis_dgemm_hn_ccc+  & 1.579 & 1.29e-08\\
\hline
\verb+blis_dgemm_hc_ccc+  & 1.575 & 1.29e-08\\
\hline
\verb+blis_dgemm_ht_ccc+  & 1.615 & 1.31e-08\\
\hline
\verb+blis_dgemm_hh_ccc+  & 1.614 & 1.28e-08\\
\hline
\multicolumn{3}{p{0.45\textwidth}}{dt=data type, op=operation, params=[(n)o-transpose, (t)ranspose, (c)onjugate, (h)ermitian-transpose]. The ``c'' and ``h'' options are the same as ``n'' and ``t'' respectively, as the tests are for real values, in this case.}
\end{tabular}
\end{table}
\pagebreak

\subsection{HPL Linpack Tests}
Finally, the High Performance Linpack Benchmark \cite{hpl} was run with the parameters and results specified in table \ref{table_hpl}. It was run with a process grid of $1\times 1$, in one node.

\begin{table}[!h]
\renewcommand{\arraystretch}{1.5}
\caption{Results for the High Performance Linpack Benchmark.}
\label{table_hpl}
\centering
\begin{tabular}{|p{0.2\textwidth}|c|c|}
\hline
N & 4608\\
\hline
NB & 768\\
\hline
P & 1\\
\hline
Q & 1\\
\hline
Time (s) & 131.81\\
\hline
GFLOPS/s & 0.495\\
\hline
$\frac{||Ax-b||_{\infty}}{(\epsilon\cdot (||A||_{\infty}\cdot ||x||_{\infty}+||b||_{\infty})*N)}$ & 21097632504.5644760\\
\hline
Residue (*) & 2.34e-06\\
\hline
\multicolumn{2}{p{0.45\textwidth}}{(*) The residue is taken as $res = \frac{||Ax-b||_{\infty}}{(||A||_{\infty}\cdot ||x||_{\infty}+||b||_{\infty})*N}$, and is calculated by multiplying the HPL result (previous row) times $\epsilon$ ($= 2^{-53}$).}
\end{tabular}
\end{table}

The results of the HPL benchmark showed that the sgemm implementation works correctly, up to Single Precision, but the performance is far lower than the one for the sgemm operation alone. The lower performance could be explained as due to a poor choice of algorithm parameters for the benchmark, or by the influence of the other BLAS functions that are called, in particular the Level-2 BLAS operations. Those Level-2 operations should not account for most of the computations, but if their performance is very low, compared to the Level-3 operations, they could be the limiting factor.

\section{Conclusion and Future Work}
An Epiphany accelerated, complete BLAS library was instantiated by the use of the BLIS framework. The performance of the Matrix-Matrix multiplication kernel achieved was better than in any other implementation before (as to the author's knowledge), when program loading and initialization are not taken into account (which is the standard in previous work \cite{varghese}\cite{ysapir}\cite{Ross2016}). When trying to get a more practical kernel, to be used as a Linux service, the performance gets lower, due to the interprocess communication (which could, most likely, be improved), but gives still an interesting result for a first BLAS implementation. The results for the High Performance Linpack are far lower than expected, given the sgemm results. That may be explained due to a poor choice of parameters for the algorithm, or to the low performance of Level-2 BLAS functions.

There are many possible improvements for this implementation. Some of them are discussed below.

\subsection{A ``b-streaming'' Solution}
One way to improve the $ir$ ratio would be to use a solution in which the values of $B$ are only copied to the local memory as needed. That solution could make use of more free space for the input $A$.

\subsection{An ``output-streaming'' Solution}
\begin{figure}[!h]
\centering
\includegraphics[width=2.5in]{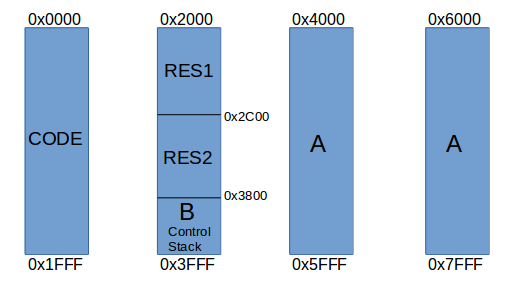}
\caption{Possible local memory mapping for one core in the ``Output-streaming'' solution. A and B are the inputs (B is not completely stored in local memory). RES1 and RES2 are used as temporary results buffers. The stack and some control variables have a reserved region. The first bank is used by the kernel's code.}
\label{fig_memoryMapStreaming}
\end{figure}

If the output is not entirely stored locally, it is possible to use bigger values for $m$ and $n$. In that kind of solutions, though, it is not possible to accumulate results for more than one $KSUB$ block, in the coprocessor. The shrinking of $RES2$, makes some more space available for the input $A$. Also it is possible to increase the value of $m$ by reducing the value of $KSUB$, but if that is done one has to make more partial results sums in the host. This idea was implemented in a previous version. Initially the idea was that summing two buffers that are stored in RAM memory would be fast enough for the host. Regretfully the access, by the host, to the shared portion of the RAM memory (HC-RAM) was very slow (at the moment it is accessed by the eSDK ``\verb+e_read+'' function), thus limiting that kind of improvements (bigger $m$,$n$ means better $ir$ ratio). It is very possible that a faster way to read from that region of the external memory exists, in which case the ``output-streaming'' solution could achieve better performance. It was found that it was possible to access the shared memory region with a normal C pointer, but the performance results were even worse than when using the standard ``eSDK'' function call. Therefore, as the access to other portions of the RAM (non-shared) is very fast, it is assumed that there is a penalty due to the hardware configuration or FPGA implementation for the shared-RAM access. The ``output-streaming'' implementation was what the author originally had in mind when implementing the ``SUMMA-like'' algorithm.\\
A posible memory map for that solution would be as in figure \ref{fig_memoryMapStreaming}.

\subsection{NEON or FPGA acceleration}
For both, the level-2 BLAS operations and the summing of partial results by the Epiphany, the NEON SIMD engine in the ARM host or the FPGA in the Zynq could be used.


\ifCLASSOPTIONcompsoc
  \section*{Acknowledgments}
\else
  \section*{Acknowledgment}
\fi

The author would like to thank to Antel for allowing and encouraging the pursuit of this line of research, to Eng. Pablo Menoni for proofreading the manuscript, to the Parallella Community \cite{parallellaForum} for their support and suggestions, and to the Adapteva team for making this exciting platform open.



\bibliographystyle{IEEEtran}
\bibliography{IEEEabrv,TasendeArXiv}
%





\vfill
\textbf{Miguel Tasende} is an Electric Engineer at the Research and Development department of Antel. His research interests include High Performance Computing, Big Data, and Artificial Intelligence. Tasende has received a MSc degree in photonics and laser techonologies from Universidad de Vigo, Spain. He has received his BSc degree in electric engineering from Universidad de la Rep\'{u}blica, Uruguay. Contact him at mtasendebracco@antel.com.uy

\end{document}